\documentclass[twocolumn,showpacs,aps,pra, floatfix,superscriptaddress]{revtex4-1}
\usepackage{amssymb,amsthm, amssymb, latexsym}
\usepackage{bbm}
\usepackage{graphicx}
\usepackage{amsmath}
\usepackage[english]{babel}
\usepackage{pstricks}
\usepackage{mathrsfs}
\usepackage{overpic}
\usepackage{CJK}

\begin{document}
\begin{CJK}{UTF8}{gkai}
\bibliographystyle{unsrt}
\title{Half-skyrmion and meron pair in spinor condensates}

\author{Yu-Xin Hu (胡渝鑫)}
\affiliation{Centre for Quantum Technologies, National University of Singapore, 3 Science Drive 2, Singapore 117543, Singapore}
\affiliation{Merlion MajuLab, CNRS-UNS-NUS-NTU International Joint Research Unit UMI 3654, Singapore}

\author{Christian Miniatura}
\affiliation{Merlion MajuLab, CNRS-UNS-NUS-NTU International Joint Research Unit UMI 3654, Singapore}
\affiliation{Centre for Quantum Technologies, National University of Singapore, 3 Science Drive 2, Singapore 117543, Singapore}
\affiliation{Department of Physics, National University of Singapore, 2 Science Drive 3, Singapore 117542, Singapore}
\affiliation{Institut Non Lin\'{e}aire de Nice, UMR 7335, UNS, CNRS; 1361 route des Lucioles, 06560 Valbonne, France}
\affiliation{Institute of Advanced Studies, Nanyang Technological University, 60 Nanyang View, Singapore 639673, Singapore}

\author{Beno\^{\i}t~Gr\'{e}maud}
\affiliation{Merlion MajuLab, CNRS-UNS-NUS-NTU International Joint Research Unit UMI 3654, Singapore}
\affiliation{Centre for Quantum Technologies, National University of Singapore, 3 Science Drive 2, Singapore 117543, Singapore}
\affiliation{Department of Physics, National University of Singapore, 2 Science Drive 3, Singapore 117542, Singapore}
\affiliation{Laboratoire Kastler Brossel, Ecole Normale Sup\'{e}rieure CNRS, UPMC; 4 Place Jussieu, 75005 Paris, France}

\date{\today}

\begin{abstract}

We propose a simple experimental scheme to generate spin textures in the ground state of interacting ultracold bosonic atoms loaded in a two-dimensional harmonic trap. Our scheme is based on 
two co-propagating Laguerre-Gauss laser beams illuminating the atoms and coupling two of their internal ground state Zeeman sublevels. Using a Gross-Pitaevskii description, we show that the ground state of the atomic system has different topological properties depending on the interaction strength and the laser beam intensity. A half-skyrmion state develops at low interactions while a meron pair develops at large interactions.

\end{abstract}


\maketitle

\section{Introduction}

Because of their ability to materialize abstract theoretical models into carefully designed and controlled experiments, ultracold quantum gases have successfully pervaded many diverse fields of physics ranging from lattice and spin systems, quantum information, quantum simulators, to gauge fields and Anderson localization to cite a few~\cite{LewensteinBook}.
This is particularly true in the condensed-matter realm where they became a key player in many-body physics as exemplified by the first observation of the Mott-superfluid transition~\cite{Lewenstein07,Blochreview08,Ketterle2}.

In recent years, the physics of the quantum Hall effects has become one important focus of the ultracold atoms community. Because atoms are neutral, one needed effective schemes to mimic the action of a magnetic field. A first idea was to set quantum gases into rapid rotation~\cite{Cooper2008} but it faded away because more promising alternatives using light-atom coupling were quickly proposed and experimentally studied~\cite{Spielman09a,Spielman09b,Spielman11,Zhang12,Zwierlein12,Windpassinger12,Spielman12}. A large variety of Hamiltonians, including non-Abelian ones, either
in lattices or in the bulk~\cite{Dalibard11,Cooper2011,Juzeliunas2012,Goldman2013}, have been now proposed to mimic magnetic field configurations like artificial Dirac monopoles~\cite{Ray2014,yuxin2014}, spin-orbit (SO) coupling~\cite{Zhu2006,Stanescu2008,Barnett2012} or topological phases~\cite{Mottonen09,Bercioux11}. For instance, for atoms loaded in a square optical lattice, SO coupling leads to highly nontrivial properties like
ground states breaking time reversal invariance and/or magnetic textures
with topological properties, like a skyrmion crystal~\cite{trivedi2012,hofstetter2012,jiri2014}. 
Such skyrmionic structures have been experimentally observed in excitations of cold atomic gases~\cite{Bigelow2009}, but not yet in the ground state. From a theorerical point of view, some papers have proposed to generate these topological configurations with cold atomic gases either in transient excitations, which decay eventually to a non-topological configuration~\cite{Zhai2003}, or directly in the ground state~\cite{Ueda2004,Ueda2005}. However the actual experimental implementation of the latter proposal remains quite challenging. 

In the present paper, we provide a rather simple experimental set-up to generate spin textures in the ground state of interacting ultracold bosonic atoms loaded in a two-dimensional harmonic trap. Our scheme is based on 
two co-propagating Laguerre-Gauss laser beams illuminating the atoms and coupling two of their internal ground state Zeeman sublevels. At the mean field level, i.e. using a Gross-Pitaevskii description, we show that the ground state of the atomic system has different topological properties depending on the interaction strength and the laser beam intensity. A half-skyrmion state, also known as a Mermin-Ho vortex~\cite{Ho1976}, develops at low interactions while a meron pair develops at large interactions.

In the following, we first introduce our model and its effective Hamiltonian, then
we briefly present the essential properties of the single particle states. Next, we analyse the topological properties
of the ground state in the weak interaction limit. Finally, we show that at large interaction there is a transition to
a ground state made of a vortex-antivortex pair separated by a finite distance. The separation between the two opposite vortices vanishes at the transition and increases with the interaction.

\section{Model Hamiltonian}

\subsection{Experimental Setup}

\begin{figure}[tb]
\centering
\includegraphics*[width=6cm]{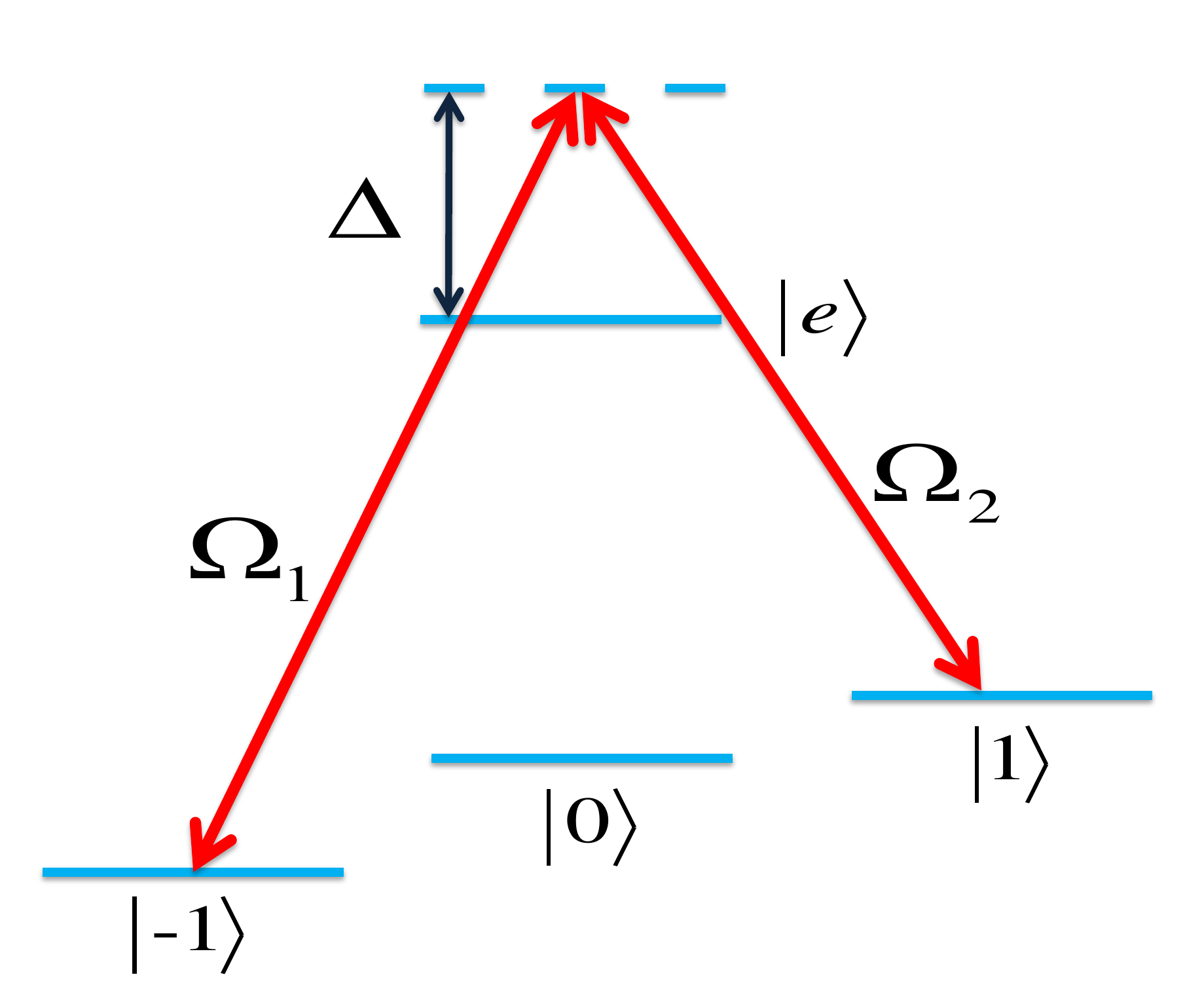}
\caption{\label{Laserconfiguration}  Atomic Zeeman diagram showing the three ground state levels $|0\rangle$ and $|\pm\!1\rangle$ splitter 
by an external static magnetic field $\boldsymbol{B}$ along axis $Oz$ (chosen as the quantization axis). 
The atoms are exposed to two far-detuned lasers beams co-propagating along $Oz$ with opposite circular polarizations. 
Only levels $m_{F}=\pm1$ are coupled to each other by resonant two-photon transitions through 
an intermediate excited state $|e\rangle$ which is non-resonantly coupled to the ground state and is thus barely populated. 
The detuning $\Delta$ is here assumed positive (blue-detuning). 
The atoms are  tightly confined in the $Oz$ direction and harmonically trapped in the transverse plane ($Ox,Oy$).}
\end{figure}

We consider here bosonic ultracold atoms with 3 internal ground state levels, for instance the $F=1$ states of $^{87}$Rb. We assume the atoms are harmonically-trapped in the two-dimensional plane ($Ox,Oy$) and tightly-confined in the third direction $Oz$ (chosen as the quantization axis) so that the atomic dynamics is effectively two-dimensional. The atoms are  further subjected to a static magnetic field along $Oz$ splitting the Zeeman degeneracy and are illuminated by two far-detuned laser beams (blue detuning) co-propagating along $Oz$ with opposite circular polarizations. 
These two laser beams create a resonant Raman coupling between the Zeeman sublevels $m_{F}=\pm1$, see Fig.~\ref{Laserconfiguration} ($\Lambda$ scheme). 
In the rotating-wave approximation, and after adiabatic elimination of the excited state, 
the effective $2 \times 2$ Hamiltonian describing the dynamics in the ($Ox,Oy$) plane for the $m_{F}=\pm1$ ground state manifold reads \cite{scullyQO}:

\begin{equation}
H_e+\left(\frac{p^2}{2m}+\frac{1}{2}m\omega^2\rho^2)\right) \, \mathbbm{1}+
\frac{1}{4\Delta}\left(    \begin{array}{cc}
            \vert \Omega_{1} \vert^{2} & \Omega_{2}\Omega_{1}^{*} \\
            \Omega_{1}\Omega_{2}^{*} & \vert \Omega_{2} \vert^{2} \\
          \end{array}
        \right)
\end{equation}
where  $m$ is the mass of the atoms, $\omega$ the harmonic trapping frequency and $r=\sqrt{x^2+y^2}$ the radial distance in the plane and where we have used the pseudo-spin representation $|\!\downarrow\rangle \equiv |m_F=-1\rangle$ and $|\!\uparrow\rangle  \equiv |m_F=1\rangle$. In the specific case of Laguerre-Gauss beams \cite{Marzlin2000,Bigelow2009} with equal (real) strength $\Omega_0$ and carrying opposite orbital angular momentum $\pm \hbar$, the respective Rabi frequencies read:

\begin{equation}
 \Omega_{1}=\Omega_{0} \, \frac{r}{R}e^{\mathrm{i}(kz+\varphi)} \quad \quad \Omega_{2}=\Omega_{0} \, \frac{r}{R} e^{\mathrm{i}(kz-\varphi)},
\end{equation}
where $R$ is the size of the "doughnut" core, $k$ the laser wave number and $\varphi$ the polar angle of vector ${\bf r}=(x,y)$. We assume here that the transverse size of the laser beams is much larger than the atomic cloud.

In the following, we use the harmonic oscillator quantum of energy $\hbar\omega$, the harmonic length $a_{ho}=\sqrt{\hbar/m\omega}$ and $\hbar/a_{ho}$ as energy, space and momentum units. We also denote the usual Pauli spin matrices by $\sigma_x$, $\sigma_y$ and $\sigma_z$. The dimensionless single-particle effective Hamiltonian then reads:
\begin{equation}
\label{H}
H_{0} = (\frac{1}{2} {\bf p}^2+\frac{1}{2}r^2) \mathbbm{1} + \frac{1}{2}\Omega^2r^2 \left(
            \begin{array}{cc}
              1 & e^{-2\mathrm{i}\varphi} \\
              e^{2\mathrm{i}\varphi} & 1 \\
            \end{array}
          \right)
\end{equation}
with $\Omega^{2}=\Omega^{2}_{0}/(2m\omega^2R^{2}\Delta)$ and where ${\bf p} = -\mathrm{i} \boldsymbol{\nabla}$. As easily checked, this Hamiltonian 
is invariant under a {\it combined} space and spin rotation, 
namely $H_0 = R(\varphi_0) H_0 R^\dag(\varphi_0)$ where $R(\varphi_0) = e^{\mathrm{i}\varphi_{0}(\hat{L}_{z}+\sigma_{z})}$ is the 
operator associated to a rotation by an angle $\varphi_{0}$ around $Oz$ both in coordinate and spin space. 
Here $L_z = -\mathrm{i} \partial/\partial\varphi$ is the orbital angular momentum operator around $Oz$. 
Applying the unitary transformation $U(\varphi) = e^{\mathrm{i}\varphi\sigma_z}$, one gets the unitary-equivalent Hamiltonian $\tilde{H}_0 = UH_{0}U^\dag$ with:
\begin{equation}\label{newH}
  \tilde{H}_0 = \frac{1}{2}({\bf p}\mathbbm{1}+\frac{\hat{e}_{\varphi}}{r}\sigma_{z})^2+\frac{1}{2}(1+\Omega^{2})r^{2}\mathbbm{1} + \frac{1}{2}\Omega^2r^2 \sigma_{x}.
\end{equation}
In this new gauge, $\tilde{H}_0$ can be viewed as the Hamiltonian 
of a particle subjected to the artificial gauge potential ${\bf A}=-\frac{1}{r}\hat{e}_{\varphi}\, \sigma_{z}$\cite{Spielman2011} 
associated to two infinite strings carrying opposite magnetic fluxes $\Phi = \pm2\pi$, one along the positive $Oz$ axis, the other one along the negative $Oz$ axis. The corresponding magnetic field is simply given by ${\bf B} = 2\pi\delta({\bf r})\hat{e}_{z} \, \sigma_{z}$. 

\subsection{Single-particle eigenstates}

Since $H_0$ is invariant under a  combined spin and space rotation, 
its spinor eigenstates in the pseudo-spin basis ($|\!\downarrow\rangle$, $|\!\uparrow\rangle$) have the general structure:
\begin{equation}
\label{singlestate}
  \underline{\phi_{m}}(\boldsymbol{r}) = 
  \left(\begin{array}{c}
  \phi_{m\uparrow}(\boldsymbol{r}) \\
  \phi_{m\downarrow}(\boldsymbol{r})
\end{array}\right)=\left(\begin{array}{c} 
  f_{m}(r)e^{-\mathrm{i}\varphi}\\
  g_{m}(r)e^{\mathrm{i}\varphi}
  \end{array} \right) \frac{e^{\mathrm{i}m\varphi}}{\sqrt{2\pi}}
\end{equation}
where $m$ is an integer. Inspection of the coupled Schr\"odinger equations for the eigenstates 
shows that both radial functions $f_{m}(r)$ and $g_{m}(r)$ can be chosen real. $H_0$ is also invariant under 
the operator $\mathcal{T}=\sigma_{x}\mathcal{C}$, $H_0 = \mathcal{T}H_0\mathcal{T}^{-1}$, 
where $\mathcal{C}$ represent complex conjugation. This implies that both $\underline{\phi_m}({\bf r})$ and $\mathcal{T}\underline{\phi_m}({\bf r})$ 
are eigenstates of $H_0$ with the same eigenenergy $\epsilon_{|m|}(\Omega)$. Since $\mathcal{T}\underline{\phi_m}({\bf r}) = \pm \, \underline{\phi_{-m}}({\bf r})$, 
we have $g_m = \pm \, f_{-m}$ and we can restrict the analysis to the $m\geq 0$ sector. 
Noting that $\underline{\phi_m}({\bf r})$ and $\mathcal{T}\underline{\phi_m}({\bf r})$ are orthogonal spinors when $m\not=0$, 
we conclude that their corresponding eigenenergy is doubly degenerate when $\Omega >0$.

\begin{figure}[htbp]
\includegraphics[width=8cm]{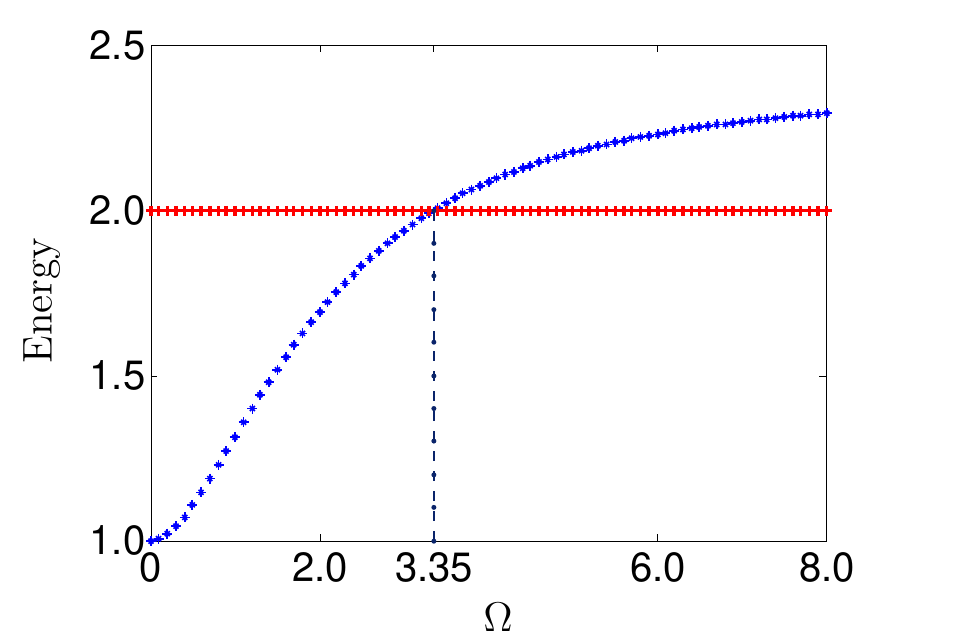}
\caption{\label{ES} The two lowest single-particle energies (in units of $\hbar\omega$) as a function of the dimensionless Rabi frequency $\Omega$.
The horizontal points (in red) correspond to the non degenerate spinor state $\underline{\phi_0}$ where $g_0=-f_0$, see text. 
The star symbols (in blue) correspond to the degenerate spinor states $\underline{\phi_{1}}$ and $\mathcal{T}\underline{\phi_{1}}$, see text.
The two energy branches cross at the critical value $\Omega_c \simeq 3.35$.
}
\end{figure}

Fig.~\ref{ES} displays the two lowest eigenenergies of $H_0$ as a function of the dimensionless Rabi frequency $\Omega$. Below $\Omega_{c}\simeq 3.35$, the ground state manifold is doubly degenerate and is spanned by the two spinor states $\underline{\phi_{1}}({\bf r})$ and $\mathcal{T}\underline{\phi_{1}}({\bf r})$.
We find that $f_1(r)$ reaches a finite value at the origin $r=0$ while $g_1(r)$ vanishes. This means that the spin-down component of $\underline{\phi_1}({\bf r})$, $g_1(r)\exp(2\mathrm{i}\varphi)$, depicts a vortex with vorticity equal to 2 while the spin-up component of $\mathcal{T}\underline{\phi_{1}}({\bf r})$, $g_1(r)\exp(-2\mathrm{i}\varphi)$, depicts the opposite vortex.
A convenient parametrization of the spinor proves to be $\underline{\phi_1}({\bf r}) = \sqrt{n_1(r)} \, \underline{\chi_1}({\bf r})$ where
\begin{equation}\label{HS}
\underline{\chi_1}({\bf r}) = \left(\begin{array}{c} \chi_{1\uparrow}({\bf r}) \\ \chi_{1\downarrow}({\bf r}) \end{array}\right)
= \left(\begin{array}{c}
  -\cos\frac{\beta(r)}{2} \\
  e^{i2\varphi}\sin\frac{\beta(r)}{2} 
\end{array}\right)
\end{equation}
and $n_1(r) =  f_1^2(r) + g_1^2(r)$ is the total density. As $n_1(r)$ is finite at the origin and $g_1(r)$ vanishes, we must have $\beta(0)=0$. 
We also find $\beta(\infty)=\pi/2$ corresponding to $n_{\uparrow}=n_{\downarrow}$ at large distances. 
A configuration satisfying such boundary conditions is known as a Mermin-Ho vortex~\cite{Ho1976}, also called a half-skyrmion since one has $\beta(\infty)=\pi$ for a "full" skyrmion. The local spin texture is defined by
\begin{equation}
\begin{aligned}
\label{MH}
&\boldsymbol{S}({\bf r}) = \underline{\chi_1}({\bf r})^\dag\boldsymbol{\sigma}\underline{\chi_1}({\bf r}) \\
&= - \sin\beta(r)(\cos2\varphi \, \hat{e}_x+\sin2\varphi \, \hat{e}_y)+\cos\beta(r) \, \hat{e}_z
\end{aligned}
\end{equation}
with modulus $|\boldsymbol{S}({\bf r})|=1$. It characterizes a 2D Skyrmion with topological charge~\cite{rajaraman1982solitons}
\begin{equation}\label{TC}
  Q = \int d^2{\bf r} \, q({\bf r}) = \int d^{2}{\bf r} \, \epsilon^{ij} \, \frac{\boldsymbol{S}\cdot (\partial_{i}\boldsymbol{S}\times\partial_{j}\boldsymbol{S}) }{8\pi}
\end{equation}
where $i,j=x,y$ and where $\epsilon^{ij}$ is the antisymmetric tensor. Using the parametrization given by Eq.\eqref{MH}, the topological charge density is simply
\begin{equation}
\label{TCD}
 q(\boldsymbol{r})=\epsilon^{ij} \, \frac{\boldsymbol{S}\cdot (\partial_{i}\boldsymbol{S}\times\partial_{j}\boldsymbol{S})}{8\pi} = -\frac{1}{2\pi r}\frac{\text{d}\cos\beta(r)}{\text{d}r}.
\end{equation}
We thus find that, for $\Omega < \Omega_c$, the topological properties of the ground state spin texture of our non-interacting system are described by a Mermin-Ho vortex with unit topological charge $Q=\cos\beta(0)-\cos\beta(\infty)=1$.

Above $\Omega_{c}$, the ground state manifold is non degenerate and the eigenstate is now the spinor $\underline{\phi_{0}}({\bf r})$ where $g_0(r) =-f_0(r)$. Since $f_0(r)$ vanishes at the origin, we see that the two spin components, $\phi_{0\uparrow}({\bf r}) = f_0(r)e^{-i\varphi}$ and $\phi_{0\downarrow}({\bf r}) =-f_0(r)e^{+i\varphi}$, describe opposite vortices with unit vorticity.

\section{Interacting bosons}

We assume here that the atoms in the $m_F =\pm 1$ Zeeman states interact through a fully $SU(2)$-symmetric interaction and are not coupled to the $m_F=0$ state. The corresponding second-quantized Hamiltonian reads
\begin{equation}
H_{int}=\frac{g}{2}\int d^2\boldsymbol{r}\, \Psi^{\dagger}_{a}\Psi^{\dagger}_{b}\Psi_{b}\Psi_{a},
\end{equation}
where $g$ is the dimensionless interaction strength and where summation over the dummy pseudo-spin indices $a$ and $b$ is understood. Here $\Psi^{\dagger}_{a}$ and $\Psi_{a}$ stand for the creation and destruction operators of a particle at point ${\bf r}$ in spin component $a = \, \uparrow,\downarrow$. They satisfy the usual bosonic commutation relations $[\Psi_{a},\Psi^{\dagger}_{b}] = \delta_{ab}$.  We next assume that, in the zero temperature limit, 
all the bosons condense into a single spinor coherent state $\underline{\Phi}({\bf r})$ with spin components $\Phi_{\uparrow}(\boldsymbol{r})$ and $\Phi_{\downarrow}(\boldsymbol{r})$
and we describe the interacting system within a mean-field approach. The Gross-Pitaevskii (GP) energy functional reads

\begin{equation}
\label{GPE}
E[g, \underline{\Phi}({\bf r})]=\int d^2{\bf r} \, \left[\underline{\Phi}^\dag H_{0} \underline{\Phi} + \frac{g}{2}  (\underline{\Phi}^\dag\underline{\Phi})^2\right]
\end{equation}
where $\underline{\Phi}^\dag\underline{\Phi} = n({\bf r}) = n_\uparrow({\bf r}) + n_\downarrow({\bf r}) = |\Phi_{\uparrow}({\bf r})|^{2}+ |\Phi_{\downarrow}({\bf r})|^{2}$ is subjected to the normalization condition $\int d^2\boldsymbol{r} \, n({\bf r}) = 1$.

\subsection{Weak interaction regime}

\subsubsection{Case $\Omega < \Omega_{c}$}

\begin{figure}[htbp]
\includegraphics[width=9cm]{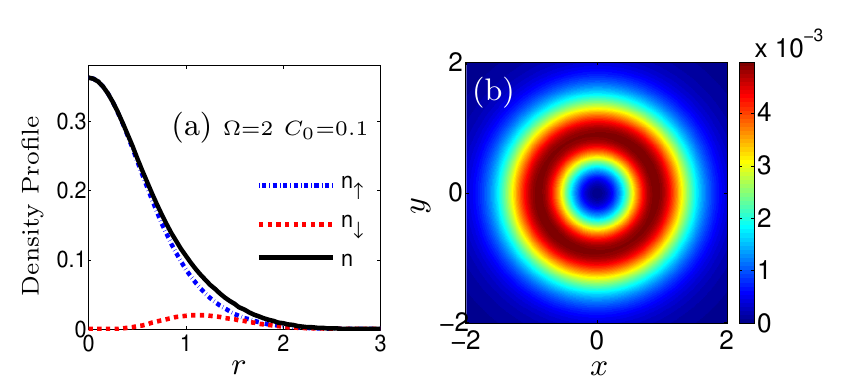}
\caption{\label{TC1} (a) Radial density profiles of the GP spinor ground state for an interaction strength $g=0.1$ and a potential energy $\Omega=2$, 
obtained from an exact numerical calculation.
(b) Corresponding topological charge density $q(\boldsymbol{r})$, see Eqs.~\eqref{MH} and \eqref{TCD}. 
Only the spin-down density $n_{\downarrow}(r)$ is vanishing at the origin, indicating that the spin-down component 
hosts a vortex with a vorticity equal to $2$. The total topological charge is $Q=\int d^2\boldsymbol{r}\,q(\boldsymbol{r})=1$.}
\end{figure}

In the limit $g\to 0$, only states with an energy separation $|\delta E|\simeq g$ or lower, are efficiently coupled. Therefore,
in first approximation, we expect the ground state $\underline{\phi_{g}}(\boldsymbol{r})$ to belong to the single-particle ground state manifold.
We thus look for the simple ansatz $\underline{\phi_{g}}(\boldsymbol{r}) =\alpha \, \underline{\phi_{1}}(\boldsymbol{r}) + \beta \,\mathcal{T}\underline{\phi_{1}}(\boldsymbol{r})$, 
where the minimization parameters $\alpha$ and $\beta$ are two constant complex numbers satisfying $\vert\alpha\vert^2+\vert\beta\vert^2=1$. 
It is easy to check that the corresponding GP energy functional is always larger than the one computed with $\underline{\phi_{1}}(\boldsymbol{r})$ 
alone (which is also equal to that computed with $\mathcal{T}\underline{\phi_{1}}(\boldsymbol{r})$ alone). This means that, when $g\to 0$, the $\mathcal{T}$-symmetry 
is spontaneously broken: $\underline{\phi_{g}}(\boldsymbol{r}) =\underline{\phi_{1}}(\boldsymbol{r})$ ($\alpha =1$) or 
$\underline{\phi_{g}}(\boldsymbol{r}) =\mathcal{T}\underline{\phi_{1}}(\boldsymbol{r})$ ($\beta=1$). 
The spin texture associated to this weakly-interacting GP ground state is a Mermin-Ho vortex with unit topological charge.
We have numerically computed the exact GP ground state and checked that the previous ansatz provides a qualitatively correct picture at small values of the interaction strength $g$. 
For instance, the density profiles $n({\bf r})$,  $n_\uparrow({\bf r})$, $n_\downarrow({\bf r})$ and the 
topological charge density $q({\bf r})$ of the exact GP spinor ground state are displayed in Fig.~\ref{TC1} for $g=0.1$ and $\Omega=2$. 
One can clearly see that $n_{\uparrow}$ remains finite whereas $n_{\downarrow}$ vanishes at the center of the trap; in addition, 
the ground state depicts a non-trivial topological charge density, with a total topological charge $Q=\int d^2\boldsymbol{r}\,q(\boldsymbol{r})=1$. 
This emphasizes that the GP ground state has the same topology as a Mermin-Ho vortex with unit topological charge.

Starting from one of the single-particle ground states selected in the limit $g\to 0$, we now increase the interaction strength $g$. 
Spinor $\underline{\phi_{0}}$ can no longer be ignored now, especially when $\Omega$ is close to $\Omega_c$ and $\underline{\phi_{1}}$, 
$\mathcal{T}\underline{\phi_{1}}(\boldsymbol{r})$ and $\underline{\phi_{0}}$ are almost degenerate. 
An updated ansatz simply reads $\underline{\phi_g}(\boldsymbol{r})=\alpha \underline{\phi_{1}}(\boldsymbol{r})+ \beta \mathcal{T}\underline{\phi_{1}}(\boldsymbol{r})+\gamma \underline{\phi_{0}}(\boldsymbol{r})$ with  
constant complex parameters satisfying $\vert\alpha\vert^2+\vert\beta\vert^2 +\vert\gamma\vert^2=1$. 
We find that there exists a critical interaction strength $g_c(\Omega)$ 
such that $\underline{\phi_g}= \underline{\phi_{1}}$ (or $\mathcal{T}\underline{\phi_{1}}$) when $g<g_c$ and  $\underline{\phi_g}=\underline{\phi_0}$ when $g>g_c$. The 
reason for this phase transition is that the spinor $\underline{\phi_0}$ 
carries less interaction energy than $\underline{\phi_{1}}$ and $\mathcal{T}\underline{\phi_{1}}$. Indeed its total density $n_0(r)$ is vanishing at the trap center whereas the total density $n_1(r)$ is maximum there. At the mean-field level, this transition is first order since the states have different vorticities.

\begin{figure}[htbp]
\centering
\includegraphics[width=6.5cm]{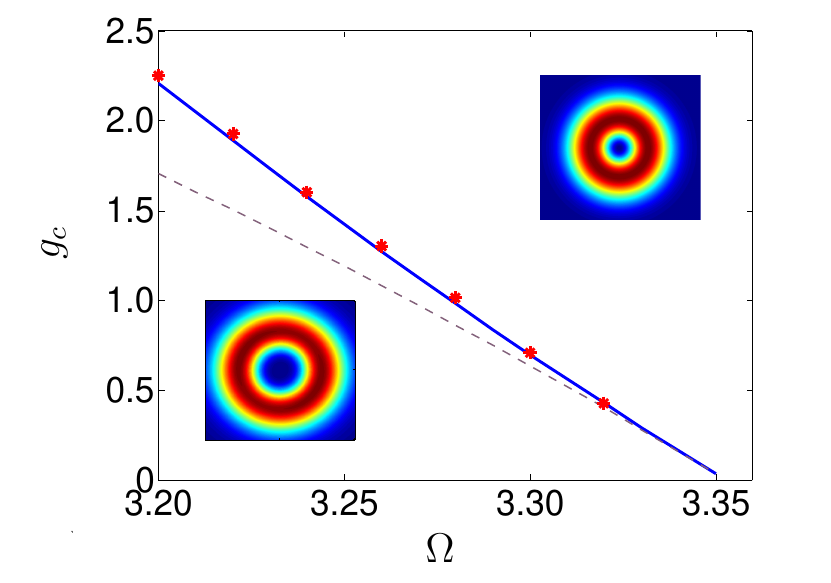}
\caption{\label{PT} Critical value $g_c$ for the transition from the $m=1$ to the $m=0$ spinor states as a function of $\Omega$. 
Dashed line: prediction obtained by comparing the GP energies computed with the single-particle spinors. 
The agreement is fair only for $\Omega$ very close to $\Omega_c$. Continuous line:  prediction obtained by comparing the GP energies 
computed with the GP ground states found, in each sector $m=0$ or $m=1$, at the interaction strength given by the dashed line. 
Star symbols: prediction obtained by minimizing the GP energy functional. 
As one can see the agreement with the continuous line is very good. The inset shows the corresponding density $n_{\uparrow}({\bf r})$ below and above the critical line.
}
\end{figure}

A rough estimate of the critical interaction strength $g_{c}$ is obtained for each $\Omega$ by equating the 
GP energy functionals computed with these single-particle ground states, namely $(\epsilon_{1}+g_{c}V_{1})$ and $(\epsilon_{0}+g_{c}V_{0})$, 
where $V_{m}=\frac{1}{4\pi}\int rdr \, n^2_m(r)$ ($m=0,1$). 
The result is shown in Fig.~\ref{PT} (dashed line). 
One may notice that this predicted $g_c$ is not really weak unless $\Omega$ is very close to $\Omega_c$. 
This means that approximating the true GP ground state by one of the single-particle states becomes questionable.
A more accurate estimate is obtained as follows. For each value of $\Omega$, we compute, in each sector (i.e. $m=0$ or $m=1$), the GP ground states 
for the interaction $g_c$ computed above. In practice, this is done by running the imaginary time evolution algorithm, starting from 
either the single-particle states $\underline{\phi_{0}}$  or $\underline{\phi_{1}}$. The invariance of the GP equation under a combined spin and space rotation ensures that the 
imaginary time evolved state always remains inside the chosen symmetry sector.
We then compute the GP functionals at interaction strength $g$ in each sector with these improved ground states 
and we find the new improved critical interaction strength by equating them. 
The result is shown in Fig.~\ref{PT} (continuous  line) and is in very good agreement with the exact value 
for $g_c$ obtained by monitoring the symmetry and the topological properties of the ground state (obtained by globally minimizing the GP energy functional)  as a function of $g$ (star symbols).

\subsubsection{$\Omega > \Omega_{c}$}

In this case, the single-particle ground state is $\underline{\phi_0}$ and 
it qualitatively describes the properties of the GP ground state in the weakly-interacting regime. This is confirmed by our
exact numerical results which show that the GP ground state indeed hosts a vortex in each of its components, but with opposite vorticity.\\

\subsection{Strong interaction regime}
\begin{figure}[htbp]
\includegraphics[width=8.5cm]{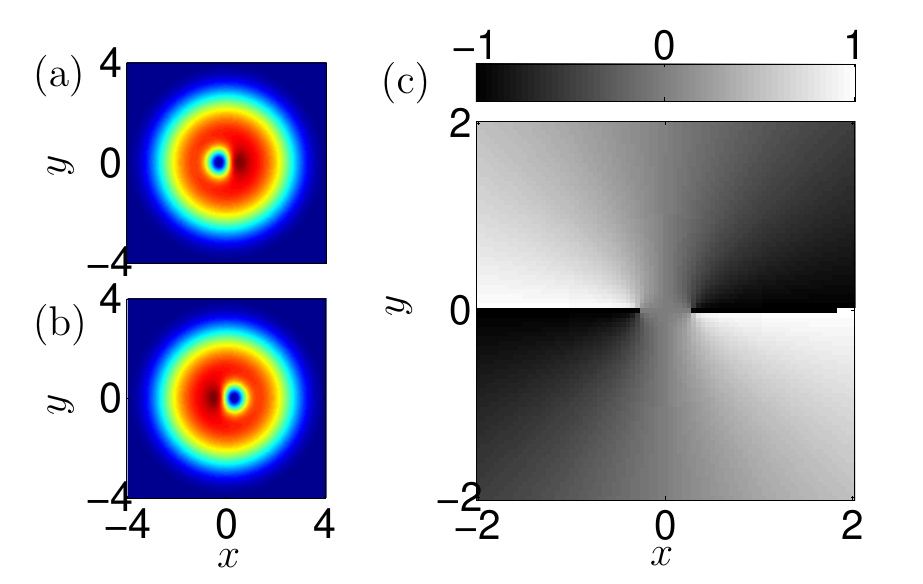}
\caption{\label{MeronDensity} Density profile of the spin-down component (a) and the spin-up component (b) of the GP ground state spinor at $\Omega=4$ and $g=100$. 
They describe a meron pair: each component hosts a vortex, the two vortices are separated by a distance $2x_m$ and have opposite vorticities. 
The relative phase $(\theta_{\downarrow}-\theta_{\uparrow})$ (in units of $\pi$) is shown in (c). There  are two clear $2\pi$ jumps on
each side of the meron pair, emphasizing that the vortex charge is $\pm 1$. One may note that the total relative phase accumulated along a loop encircling the two vortices is $4\pi$, a situation that differs from the usual meron pair~\cite{Ueda2004} where the phase jump happens when one crosses the line connecting the vortex centers.}
\end{figure}

In the strong interacting regime, higher single-particle states are coupled and no simple ansatz can be made. 
In this case we obtain the interacting ground state $\underline{\phi_g}$ by direct minimization of the GP energy functional Eq.~\eqref{GPE}. 
This is achieved by imaginary-time evolution of the corresponding GP equation. Fig.~\ref{MeronDensity} shows the ground state density of the up 
and down components and their relative phase $(\theta_\downarrow-\theta_\uparrow)$ when $g=100$ and $\Omega =4$.
As one can see, each component density vanishes at an off-centered location, at which the other component reaches its maximum, reducing thereby 
the overlapping area between the two components and therefore their interaction energy. The two points at which the densities vanish are 
located at symmetric positions $x=\pm x_m$ with respect to the center of the trap. In addition, the relative phase between the two components exhibits two clear $2\pi$ jumps
along the two segments $]-\infty,-x_m]$ and $[x_m,+\infty[$ on axis $Ox$.

\begin{figure}[htbp]
\includegraphics[width=8cm]{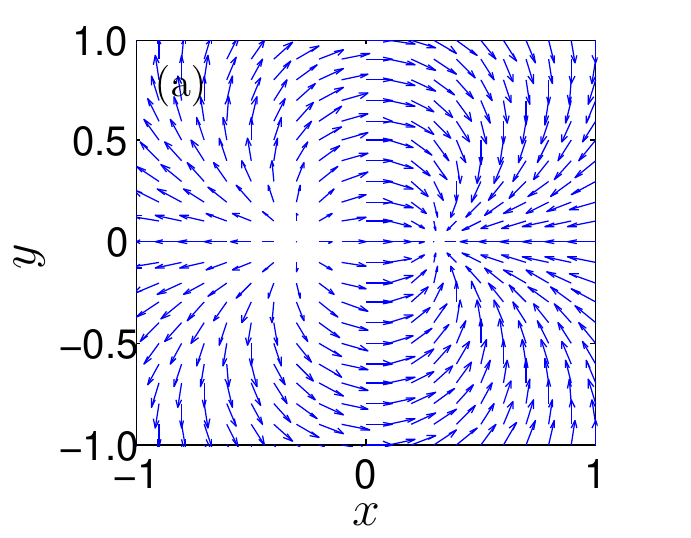}\\
\includegraphics[width=8cm]{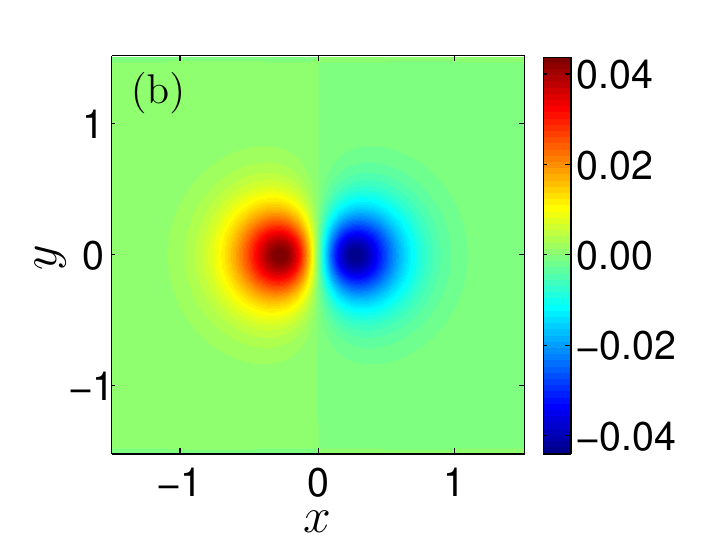}
\caption{\label{ST} Topological properties of the GP ground state for $\Omega=4$ and $g=100$. 
(a) Spin texture components $(S_x,S_y)$ in the plane $(Ox,Oy)$. 
Around the left vortex $\boldsymbol{S}=-\mathbf{e}_z$, the texture corresponds to $(S_{x},S_{y})\propto (\delta x,\delta y)$, 
whereas it corresponds to $(S_{x},S_{y}) \propto (-\delta x,-\delta y)$ around the right vortex $\boldsymbol{S}=\mathbf{e}_z$, see text. 
(b) The corresponding topological charge density, see Eq.\eqref{TC} and Eq.~\eqref{TCmeron}. T
he distribution is odd with respect to the $x$ coordinate, which emphasizes the creation of a vortex-antivortex pair with vanishing total topological charge.
}
\end{figure}

To gain further insight, we introduce again the pseudo-spin representation and decompose the GP spinor components as $\phi_{g,a}=\sqrt{n} \, \chi_{a}$ with $\chi_{a}=\vert\chi_{a}\vert e^{i\theta_{a}}$ ($a=\uparrow,\downarrow$). The total density is $n = \vert\phi_{\uparrow}\vert^2 + \vert\phi_{\downarrow}\vert^2$ and the spinor $\underline{\chi}$ thus satisfies $\vert\chi_{\uparrow}\vert^2+\vert\chi_{\downarrow}\vert^2=1$.
The corresponding local spin texture $\boldsymbol{S}=\underline{\chi}^\dag\boldsymbol{\sigma}\underline{\chi}$ reads
\begin{equation}
\begin{aligned}
&S_{x}=2\vert\chi_{\uparrow}\vert\vert\chi_{\downarrow}\vert\cos(\theta_{\downarrow}-\theta_{\uparrow})\\
&S_{y}=2\vert\chi_{\uparrow}\vert\vert\chi_{\downarrow}\vert\sin(\theta_{\downarrow}-\theta_{\uparrow})\\
&S_{z}=\vert\chi_{\uparrow}\vert^2-\vert\chi_{\downarrow}\vert^2.
\end{aligned}
\end{equation}
and has unit modulus $\vert \boldsymbol{S}\vert=1$. 
This local spin is parallel to axis $Oz$, namely $\boldsymbol{S}=\hat{ e}_z$ (resp. $\boldsymbol{S}=-\hat{ e}_z$) at space points where $n_{\downarrow}$  (resp. $n_{\uparrow}$) vanishes. 
The relative phase $(\theta_{\downarrow}-\theta_{\uparrow})$ is undefined at these two points and they correspond to a vortex-antivortex pair. These properties
appear clearly in Fig.~\ref{ST}a where the spin components $(S_{x},S_{y})$ are plotted in the plane $(Ox,Oy)$. Writing ${\bf r}=(-x_m+\delta x,\delta y)$, we find $(S_{x},S_{y})\propto (\delta x,\delta y)$ around the left vortex $\boldsymbol{S}=-\hat{e}_z$. By the same token, writing ${\bf r}=(x_m+\delta x,\delta y)$, we find $(S_{x},S_{y}) \propto (-\delta x,-\delta y)$ around the
right vortex $\boldsymbol{S}=+\mathbf{e}_z$. The topological charge density, computed with Eq.~\eqref{TC}, is displayed in Fig.~\ref{ST}. 
It emphasizes that the GP ground state spinor depicts two vortices with opposite vorticity (with respect to $S_z$), such that the total topological charge vanishes. 
 
In the pseudospin representation~\cite{Ueda2005PRA,Ueda2005}, the GP energy functional reads
\begin{equation}\label{pseudospin}
\begin{aligned}
  E=&\int d^{2}r \bigg[ \frac{1}{2}(\boldsymbol{\nabla}\sqrt{n})^2+\frac{n}{8}(\boldsymbol{\nabla}\boldsymbol{S})^2
  +\frac{n}{2}\boldsymbol{v}_{e}^2+\frac{r^2}{2}n\\
  &+\frac{1}{2}\Omega^{2}r^2(I+S_{x}\cos2\varphi+S_{y}\sin2\varphi)+\frac{g}{2}n^{2}\bigg],
\end{aligned}
\end{equation}
where 
\begin{equation}
(\boldsymbol{\nabla}\boldsymbol{S})^2 \overset{\text{def}}{=} (\boldsymbol{\nabla}S_x)^2+(\boldsymbol{\nabla}S_y)^2+(\boldsymbol{\nabla}S_z)^2. 
\end{equation}
The effective velocity field is given by~\cite{Ueda2005PRA,Ueda2005}
\begin{equation}
\begin{aligned}
\boldsymbol{v}_{e}&=\frac{1}{2}\big[\boldsymbol{\nabla}\Theta+\frac{S_{z}(S_{y}\boldsymbol{\nabla} S_{x}-S_{x}\boldsymbol{\nabla} S_{y})}{S_{x}^{2}+S_{y}^{2}} \big],
\end{aligned}
\end{equation}
and depends on the gradient of the total phase $\Theta=\theta_{\uparrow}+\theta_{\downarrow}$ and of the pseudo-spin.
In analogy with the meron pair solution discussed in \cite{Sinova2000,Ueda2004,Takao2005half}, we parameterize the spin texture as follows:
\begin{equation}\label{Meron}
\begin{aligned}
 &S_{x}=\frac{-r^2\cos 2\varphi + \lambda^2 e^{-\alpha r^2}}{r^2+\lambda^2 e^{-\alpha r^2}} \quad  S_{y}=\frac{-r^2\sin 2\varphi}{r^2+\lambda^2 e^{-\alpha r^2}}\\
 &S_{z}=-\frac{2 \lambda e^{-\alpha r^2/2} \, r\cos \varphi}{r^2+\lambda^2 e^{-\alpha r^2}}.
\end{aligned}
\end{equation}
The usual meron pair parametrization is obtained for $\alpha=0$. The corresponding topological charge density is
\begin{equation}
 \label{TCmeron}
 q(\boldsymbol{r})=-\frac{\mu \, x}{\pi\left(r^2+\mu^2\right)^2}
 -\alpha\frac{\mu \, x r^2}{\pi\left(r^2+\mu^2\right)^2},
\end{equation}
where $\mu=\lambda e^{-\alpha r^2/2}$.
The vortex-antivortex nature of the meron pair results in  a topological  
density $q(\boldsymbol{r})$ which is an odd function of coordinate $x$, see Fig.~\ref{ST}b. As a consequence, the total topological charge is $Q=\int d\boldsymbol{r}\,q(\boldsymbol{r})=0$.

The spin texture Eq.~\eqref{Meron} corresponds to the GP spinor condensate:
\begin{equation}
 \phi_{\uparrow}=\sqrt{\frac{n}{2}} \ \frac{\mu-re^{-i\varphi}}{\sqrt{r^2+\mu^2}} \qquad
 \phi_{\downarrow}=\sqrt{\frac{n}{2}} \ \frac{\mu+re^{i\varphi}}{\sqrt{r^2+\mu^2}},
\end{equation}
The meron pair is polarized along axis $Ox$ due to the $\sigma_{x}$-term in Eq.~\eqref{newH} which describes an effective magnetic field along $Ox$. 
The locations of the two vortex cores are determined by the two extremas of $S_{z}$. They are found at $(\pm x_m,0)$ where \cite{Ueda2004}
\begin{equation}
  x_{m}^2=\lambda^2 e^{-\alpha x_{m}^2}.
\end{equation}
The relative phase is given by
\begin{equation}
\label{RelPhase}
e^{i(\theta_\downarrow-\theta_\uparrow)}=\frac{\mu^2-r^2e^{2i\varphi}}{\sqrt{(r^2+\mu^2)^2-4\mu^2r^2\cos^2{\varphi}}}
\end{equation}
and is singular at the two vortex cores $(\pm x_m,0)$. Writing $(x,y)=(\pm x_m+\delta x,\delta y)$, 
a first-order expansion gives $\theta_\downarrow-\theta_\uparrow=\delta\varphi+\pi$ around $(x_m,0)$ and $\theta_\downarrow-\theta_\uparrow=\delta\varphi$ 
around $(-x_m,0)$, where $\delta\varphi=\arctan{(\delta y/\delta x)}$ is the local polar angle.
When circling around each vortex core, the accumulated relative phase is $2\pi$. 
Similarly, in the large distance limit $r\gg x_m$, the relative phase is
$\theta_\downarrow-\theta_\uparrow=2\varphi+\pi$ and a full loop around the two vortices generates a total phase change of $4\pi$. This is slightly different from the usual meron pair situation \cite{Ueda2004}, where the relative phase reaches a constant value at large distance, 
which corresponds to a spin texture pointing in a fixed direction. In the present case, $(S_x,S_y)\approx(-\cos{2\varphi},-\sin{2\varphi})$.
This difference explains why, in the present situation, the phase jumps happen on each outer side of the meron pair and not in between the two vortices, see Fig.~\ref{MeronDensity}b. Apart from this, 
the GP ground state properties are similar to those of the meron pair already studied in a double-layer quantum Hall  system \cite{Ueda2004,girvin1999quantum}. 

\begin{figure}[htbp]
\includegraphics[width=8cm]{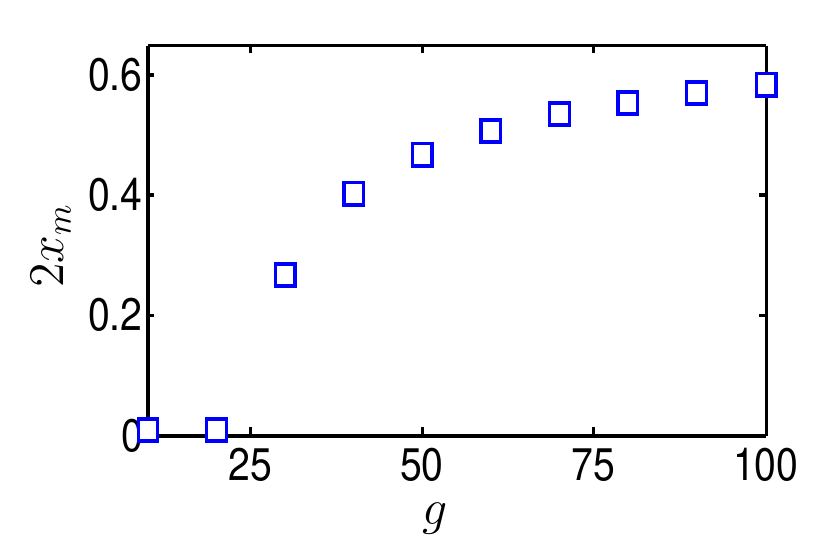}
\caption{\label{PMP} The size $2x_m$ of the meron pair as a function of $g$ for $\Omega=4$. 
The system exhibits a second-order phase transition at $g\approx 20$ between a ground state with topological properties similar to the $m=0$ 
single-particle state and a meron pair with two opposite and off-centered vortices.}
\end{figure}

By fitting our numerical data with ansatz Eq.~\eqref{Meron}, we have determined the parameters $\lambda$ and $\alpha$ as a function of $\Omega$ and $g$. The results are shown in Fig.~\ref{PMP} for $\Omega=4$. 
One can clearly see a phase transition happening at $g\approx 20$. Below, the GP ground state exhibits topological properties similar to the $m=0$ single-particle state. Above, the GP ground state describes a meron pair with two off-centered and opposite vortices. It means that the energy cost to separate and shift away the vortex cores is less than the interaction energy. Above the transition point, 
the value of $\lambda$ increases with $g$, which means that the size $2x_{m}$ of the meron pair increases. Finally, from the pseudo-spin point of view, the transition occurs between a uniformly vanishing $S_z({\bf r})$ component and a well-defined structure $S_z({\bf r})$. Therefore we expect the spin susceptibility along axis $Oz$ to diverge at the transition and  the phase transition is second order.

\section{Conclusion}
In this paper we have proposed an experimental scheme leading to non-trivial spin textures in
the interacting ground state of a two-component spinor condensate. More precisely, we have shown that a second-order phase transition occurs between a Mermin-Ho vortex and a meron pair when the interaction strength increases. A possible extension of the work is to study the excitations of the system and their topological properties. Finally, from an experimental point of view,  $F=1$ spinor condensates have also an effective spin-spin interaction $g_2 S_z^2/2$.
This interaction term breaks the $SU(2)$ invariance and converts a pair of bosons in the $|m_F=-1\rangle$ and $|m_F=+1\rangle$ spin states into a pair of bosons in the $|m_F=0\rangle$ spin state. In the present situation,
these collision processes correspond to losses. In the case of $^{87}$Rb, fortunately $g_2\ll g$ and one should be able to observe the Mermin-Ho vortex and the meron pair before
the effect of spin-spin interaction sets in. An alternative would be to lift the energy of the $|m_F=0\rangle$ spin state
and suppress the detrimental pair conversion processes by rendering them energetically less favorable.

\medskip
The Centre for Quantum Technologies is a Research Centre of Excellence funded by
the Ministry of Education and National Research Foundation of Singapore.

\end{CJK}

\end{document}